\documentclass[preprint,11pt]{aastex}

\shorttitle{NGC 2264 Kinematics}
\shortauthors{Tobin et al.}

\newcommand{\msun}{\mbox{$M_{\sun}$}}

\newcommand{\kms}{\mbox{km s$^{-1}$}}
\newcommand\etal{{\it et al}. }

\begin{document}

\title{Kinematic and Spatial Substructure in NGC 2264\footnotemark}
\author{John J. Tobin\altaffilmark{2,5}, Lee Hartmann\altaffilmark{3}, 
Gabor F{\H u}r{\'e}sz\altaffilmark{4}, Wen-Hsin Hsu\altaffilmark{3}, Mario Mateo\altaffilmark{3}}

\begin{abstract}
We present an expanded kinematic study of the young cluster NGC 2264
based upon optical radial velocities measured using multi-fiber echelle
spectroscopy at the 6.5 meter MMT and Magellan telescopes.
We report radial velocities for 695 stars, of which approximately 407 stars are
confirmed or very likely members. Our results more than double the number of
members with radial velocities from F{\H u}r{\'e}sz \etal, resulting in a much better defined
kinematic relationship
between the stellar population and the associated molecular gas.
  In particular, we find that there is a significant
subset of stars that are systematically blueshifted with respect to the
molecular ($^{13}$CO) gas.  The detection of Lithium absorption and/or
infrared excesses in
this blue-shifted population suggests that at least some of these stars are cluster members;
we suggest some speculative scenarios to explain their 
kinematics.  Our results also more clearly define the redshifted
population of stars in the northern end of the cluster; we suggest that
the stellar and gas kinematics of this region are the result of a bubble
driven by the wind from O7 star S Mon.  Our results emphasize the complexity
of the spatial and kinematic structure of NGC 2264, important for eventually
building up a comprehensive picture of cluster formation.
\end{abstract}

\footnotetext[1]{Observations reported here were obtained at the MMT Observatory, a joint facility of the
Smithsonian Institution and the University of Arizona;
 This paper includes data gathered with the 6.5 meter Magellan Telescopes
 located at Las Campanas Observatory, Chile}
\altaffiltext{2}{Hubble Fellow, National Radio Astronomy Observatory, Charlottesville, VA 22903}
\altaffiltext{3}{Department of Astronomy, University of Michigan, Ann Arbor, MI 48109}
\altaffiltext{4}{Center for Astrophysics, 60 Garden Street, Cambridge, MA 02138}
\altaffiltext{5}{Current Address: Leiden Observatory, Leiden University, P.O. Box 9513, 2300-RA Leiden, The Netherlands; tobin@strw.leidenuniv.nl}

\section{Introduction}

Most stars form in clusters \citep{carpenter2000, ladalada2003, allen2007,krumholz2014}, and
it is important to understand the processes that lead to the formation of these clusters, in 
addition to physics of the formation of the individual stars themselves.
Star cluster formation theories are generally comprised of two classes: 
highly dynamic, non-equilibrium models that form within a crossing time \citep[e.g.][]{bonnell2003,bate2012}
and quasi-equilibrium and/or slow contraction scenarios that require several crossing
times to form \citep[e.g.,][]{tkm2006}. Studies of both the spatial distribution of stars \citep[e.g.,][]{feigelson2013}
and their kinematics \citep[e.g.,][]{furesz2006, furesz2008,tobin2009,cottaar2014,foster2014} 
in young clusters that may not be dynamically relaxed and still have
substantial mass in molecular gas can help distinguish
between these two main possibilities.

The closest region of significant clustered star formation after the Orion Nebula Cluster
is NGC 2264 (see \citealt{dahm2008}) at a distance of $\sim$ 760 - 900 pc \citep{sung1997, baxter2009}.
This cluster is spatially elongated along a distance $\sim 8$~pc with significant sub-clusterings.
Combining mid-infrared data from the {\em Spitzer Space Telescope} with previous
optical photometric and H$\alpha$ emission surveys, \citet{sung2009} identified several 
distinct regions: a large, extended region of low extinction $\sim 3.5$~pc in diameter centered 
on the O7V $+$ B1.5V binary S Mon \citep{skiff2013}.
The two other sub-clusters are the ``Cone'' and ``Spokes''
(Texeira et al. 2006), and they are denser and more highly extincted, each centered around very luminous protostars.
\citet{sung2008} and \citet{feigelson2013} also highlighted a ``halo'' population of young stars that were distributed
throughout the cluster.

\citet{furesz2006} (Paper I) found a strong correlation between the velocities of
the molecular gas and the associated stellar population as a function of position,
with the optically-visible members of the subclusters showing distinct velocity components.
The correlation of spatial substructure with kinematics clearly demonstrates that NGC 2264 is
not dynamically relaxed, consistent with the youth of the stellar population
\citep{sung2008,sung2009}.  In this paper we present new radial velocity observations which,
when added to the previous measurements over the past six years, enable a refined study of the
spatio-kinematic structure of the cluster.  We defer a characterization of the
spectroscopic binary population to a subsequent paper.
While the overall correlation between stellar and gas motions remains much the same as in \citet{furesz2006},
we find additional velocity substructure that was not apparent in our initial study.
The results contribute to the development of a quantitative picture of the kinematics of NGC 2264
which can ultimately be used as a testing ground for theories of cluster formation.

\section{Observations and Data Reduction}
\label{sec:obs}

\subsection{Target Selection}
\label{sec:targets}

Because of the proximity of NGC 2264 to the Galactic Plane, careful
selection of candidate members was necessary to reduce contamination
as much as possible. This continuing study uses the same target selection
from \citet{furesz2006} which drew sources from X-ray \citep{ramirez2004},
H$\alpha$ surveys, and some additional sources showing UV or IR excess from
\citet{park2000}. We have added additional targets from \citet{rebull2002}
whose V-band magnitudes
fall within the range of magnitudes expected for cluster members.
Selection from these catalogs attempts to identify pre-main sequence cluster members in different
stages of evolution; stars from the H$\alpha$ and \citet{park2000} optical study
will preferentially find Classical T Tauri Stars (CTTSs) or Class II accreting
objects while the X-ray sources will find Class III or weak T Tauri Stars (WTTSs) without
substantial accretion activity. The additional targets from \citet{rebull2002}
are meant to fill unused fibers, enabling us to identify additional members
not selected based on other criteria.

A subset of stars whose membership was questionable because their radial velocities
differ significantly from the molecular gas in the region 
(the ``blueshifted population''; section 3 and 4.2) were selected as the main targets for
multi-slit observations with the Inamori-Magellan Areal Camera \& Spectrograph (IMACS; Bigelow \& Dressler 2003) (Section \ref{sec:imacs}) to verify membership via detection of Li I absorption.

\subsection{Hectochelle}
\label{sec:hecto}

We have observed NGC 2264 with the multi-fiber echelle spectrograph
Hectochelle \citep{szent1998} on the MMT during several epochs over the past
six years.  Hectochelle uses robots to position 240 fibers on a 1$^{\circ}$ field of view (FOV).
The fibers subtend 1\farcs4 on the sky
yielding a spectral resolution of $R \sim$ 35000.
There are limitations of fiber spacing in a given configuration;
no targets may be closer than 30$^{\prime\prime}$.
We used an order-separating filter (RV31) to cover the wavelength range of $\sim$5150 - 5300\AA.

The spectra were reduced
with an automated pipeline developed by G. F{\H u}r{\'e}sz that utilizes standard spectral
 reduction procedures within the Image Reduction and
Analysis Facility (IRAF)\footnote{IRAF is distributed by the National Optical Astronomy Observatories,
which are operated by the Association of Universities for Research
in Astronomy, Inc., under cooperative agreement with the National
Science Foundation.}. A more detailed description
of Hectochelle data reduction can be found in \citet{aurora2006}.
An additional manual step not taken care of by the pipeline was sky
subtraction. A number of fibers ($\sim$20) must be allocated for sky observations because
Hectochelle observations are generally conducted in bright time and the
scattered moonlight must be subtracted to measure velocities for faint sources.
To subtract the sky observations, we normalized the fiber
throughputs using the flat-field exposure, and then subtracted the
average spectrum taken from all the sky fibers.

\citet{furesz2006} observed NGC 2264 in March 2004 and December 2005;
our new observations were conducted in the fall of 2007, 2008, spring 2009, fall 2009, and
spring 2010, see Table 1 for details of the observations. When possible two epochs
were taken of each field in a particular observing season to identify short period 
radial velocity variability. Because Hectochelle
observations are conducted in queue mode, the second epochs could be days or
even a month apart from the first epoch.

\subsection{MIKE Fibers}
\label{sec:mmfs}

We also used MIKE Fibers \citep{bernstein2003,walker2007} on the Magellan Clay telescope to
observe NGC 2264 in the spectral range $\sim$5150 - 5210\AA.
This instrument uses 256 manually-plugged fibers that fit within a pre-drilled plate
on a 25$^{\prime}$ FOV. Each fiber subtends 1\farcs25 on the sky, resulting in a resolution
$R \sim$ 18000 and fibers must be spaced at least 14$^{\prime\prime}$ apart.
MIKE Fibers has two independent spectrograph channels;
at 5200\AA\ we are able to use both channels with 128 fibers going to
each channel.  The two channels are essentially separate
spectrographs with different gratings, optics, and CCDs in the
same enclosure.

Because MIKE Fibers has a smaller field of view than Hectochelle, we selected two
fields with the highest stellar density for these observations: one in the northern part of the cluster
near the O7V star S Mon, and another in the southern part centered between
the Spokes and Cone ``clusters''.  As in the case of the Hectochelle observations, data were taken on
two epochs several days apart.

The data were reduced using IRAF. We used the task \textit{ccdproc}, to carry out initial
reductions on the raw CCD imaging data:
subtracting the overscan, trimming, and subtracting the bias. We combined our individual
exposures with `imcombine' set for cosmic ray rejection.
The spectra were extracted using `twodspec', by first tracing
the apertures in a flat field frame taken of a continuum source to create a map of
aperture traces to extract the science spectra. The wavelength solution
was calibrated by fitting a 4th order Legendre polynomial to the
Thorium-Argon (Th-Ar) lamp spectrum taken before and after the sets of science observations.

\subsection{IMACS spectra}
\label{sec:imacs}

Low-resolution spectra for the purpose of determining membership from Li I 6707 \AA\ absorption
were obtained with IMACS on the Magellan Baade telescope, during December 20-22 2010 (UT).  
The setup was the same as used in our study of the L1641 region of the
Orion A cloud \citep{hsu2012}.  We used the IMACS f/2 camera in multi-slit spectroscopy mode with the 
300 line grism at a blaze angle of 17.5$^{\circ}$. With a 0\farcs6 slit, this configuration yields a 
resolution of 4 \AA\ and spectral coverage from approximately 4000 \AA\ to 9000 \AA. 
The standard observation time for each field is 5 x 10 minutes, but we increased the time to 6 x 10 minutes for a few observations at higher airmasses.  A total of 160 stars were observed, selected
primarily from the stars that appear slightly blueshifted from the gas (see Section 3 and 4.2).

\subsection{Radial velocity measurements}
\label{sec:rv}

We used \textit{rvsao} package in IRAF \citep{mink1998}, to measure the radial velocities of our
observed targets. The radial velocity of an object is determined by cross-correlating the observed
spectrum with a template spectrum of known velocity.
The cross-correlation signal-to-noise (S/N) or quality 
is given by $R$, defined as
\begin{equation}
R = \frac{h}{2^{1/2} \sigma_a}
\end{equation}
where $h$ is the peak height of the correlation function
and $\sigma_a$ is the rms noise estimated from the antisymmetric
portion of the correlation function. The velocity measurement error also depends
on the spectral resolution; it is of the form
\begin{equation}
\sigma_v \sim \frac{C}{1+R}
\end{equation}
where $C$ is 20 \kms\ for MIKE Fibers and 14 \kms\ for Hectochelle. These values
were determined empirically by adding random noise to a high S/N spectrum as in \citet{hartmann1986}.

We used libraries of synthetic stellar spectra as velocity templates
rather than observed velocity standards, enabling us to explore a
wider range of stellar parameters than a few observed templates.
We used the library by \citet{coelho2005} for velocity measurement
of spectra from both Hectochelle and MIKE Fibers. The \citet{coelho2005}
templates are computed with a resolution of R $\sim$ 100,000, much higher resolution than the
observed spectra.
However, a cosine-bell filter function is applied to the Fourier transform of the template and observed  
spectrum within the \textit{xcsao} task. This effectively reduces
the template resolution to that of the Hectochelle or MIKE Fibers instrument. The filter applied
to the data from both instruments (and templates) has an inner cutoff at a wave number of 10 and the 
filter function increases to 1 at a wave number of 40. The purpose of the inner
cutoff is to limit the effect of large-scale features in the cross-correlation. 
The outer cutoff values depend on the instrument and limit
on the smallest features considered in the cross-correlation. \citet{tonry1979} suggests 
an outer cutoff of $\sim$ N$_{pixels}$/(2$\pi$ $\times$ FWHM/2.355), yielding
$\sim$400 for Hectochelle and $\sim$80 for MIKE Fibers. However,
we found that outer wavenumber cutoff beginning at 120 and reaching zero at 450
worked the best for MIKE Fibers and a cutoff beginning at 600, reaching zero at 1000 worked best for
Hectochelle.

We used a subset of the spectral templates with surface gravity log(g) = 3.5,
effective temperatures (T$_{eff}$) ranging from 3500 - 7000K in steps of 250K, and 
solar metallicity. There was no need to explore a wider parameter space
with the templates because the most important factor determining the
quality of a correlation was the T$_{eff}$ of the template.

The results from the correlation with the highest $R$ value
are selected, matched to the appropriate target coordinates, and stored in a
Starbase database \citep{roll1996}. The right ascension and declination 
are used to correlate the new observations with the existing catalog
and the coordinates are used to generate a truncated 2MASS ID number (2MASSID), throwing out
fractional seconds in right ascension and declination. Once the targets have been
matched in right ascension and declination, the 2MASSID is used to identify sources
in further analysis.

To combine radial velocities taken in different epochs, it is necessary to determine zeropoint
velocity shifts for an entire observation. This compensates for possibly different 
calibration schemes, different wavelengths, different instruments, as well as temperature
variations within the spectrograph \citep{aurora2006,furesz2006,furesz2008}. 
We also accounted for the well-characterized
fiber-to-fiber velocity offset caused by the calibration
lamps not illuminating the fibers in the same way as astronomical objects.

We adopted the results of \citet{furesz2006} as baseline velocities to shift our observations to match. 
This is because the \citet{furesz2006} velocities are referenced to the radial velocity standard star W23870 \citep{latham1991}, having
an absolute velocity accuracy of $\pm$0.2 \kms\ \citep{stefanik1999}.
We applied a constant shift for each field given by
the mean of a Gaussian fit to the distribution of radial velocity differences for each target in a field.
The zeropoint shifts for the Hectochelle and MIKE Fibers fields are generally quite small, less than 0.5 km s$^{-1}$
and are given in Tables 1 and 2. In contrast, observations of the ONC in \citet{tobin2009} had zeropoint
shifts $\sim$1 km s$^{-1}$. Given than many NGC 2264 fields were conducted in the same semester
and even same nights as the Orion fields, the zeropoint shifts may be overestimated in
\citet{tobin2009}. We believe that the NGC 2264 fields track the zeropoint shift better than the
Orion fields due to the inclusion of many bright non-members in the Galactic Plane that give velocity
precisions of 0.5-0.2 \kms\ (note that this is the formal error and does not include systematic effects.)

We have converted the heliocentric radial velocities of each target to the
kinematic local standard of rest (LSR) velocities \citep{kerr1986}. 
In order to compare the velocity structure of the stars and gas, 
the LSR velocities are used exclusively in the plots, but the heliocentric radial velocities 
are given in Table 3. 

The multi-epoch observations have enabled us to also identify a number of candidate spectroscopic binaries
from radial velocity variability and/or double-peaked correlation functions.
We have removed these from the cluster kinematic analysis because they will broaden the
cluster velocity distribution. These candidate spectroscopic binaries will be presented in 
a subsequent paper along with observations from later epochs.

\subsection{Li and spectral types}
\label{sec:li.sp}
Spectral types and Li I 6707 \AA\ absorption equivalent widths
were determined for a subset of target stars observed with IMACS.
We used SPTCLASS, a semi-automatic spectral-typing program 
\citep{hernandez04}, which uses empirical relations of 
spectral type and absorption line equivalent widths for
classification (see \citet{hsu2012} for details).
SPTCLASS also automatically measures Li I equivalent widths, but we made
manual determinations because of the uncertainty in automatically
setting continuum levels at this resolution when strong TiO bands are present.

\section{Results}

The combined radial velocity measurements from
our observations are presented in Table 3. Of the 695 stars with
measured non-variable radial velocities, approximately 407 are probable members based
on their radial velocities, with more certain identification for those objects
with IR excess and/or Li I absorption.  This approximately doubles the number of
stars with radial velocities (excluding spectroscopic binaries) available in the
previous study of (Paper I). 

\begin{figure}
\begin{center}
\includegraphics[scale=0.75, angle=-90]{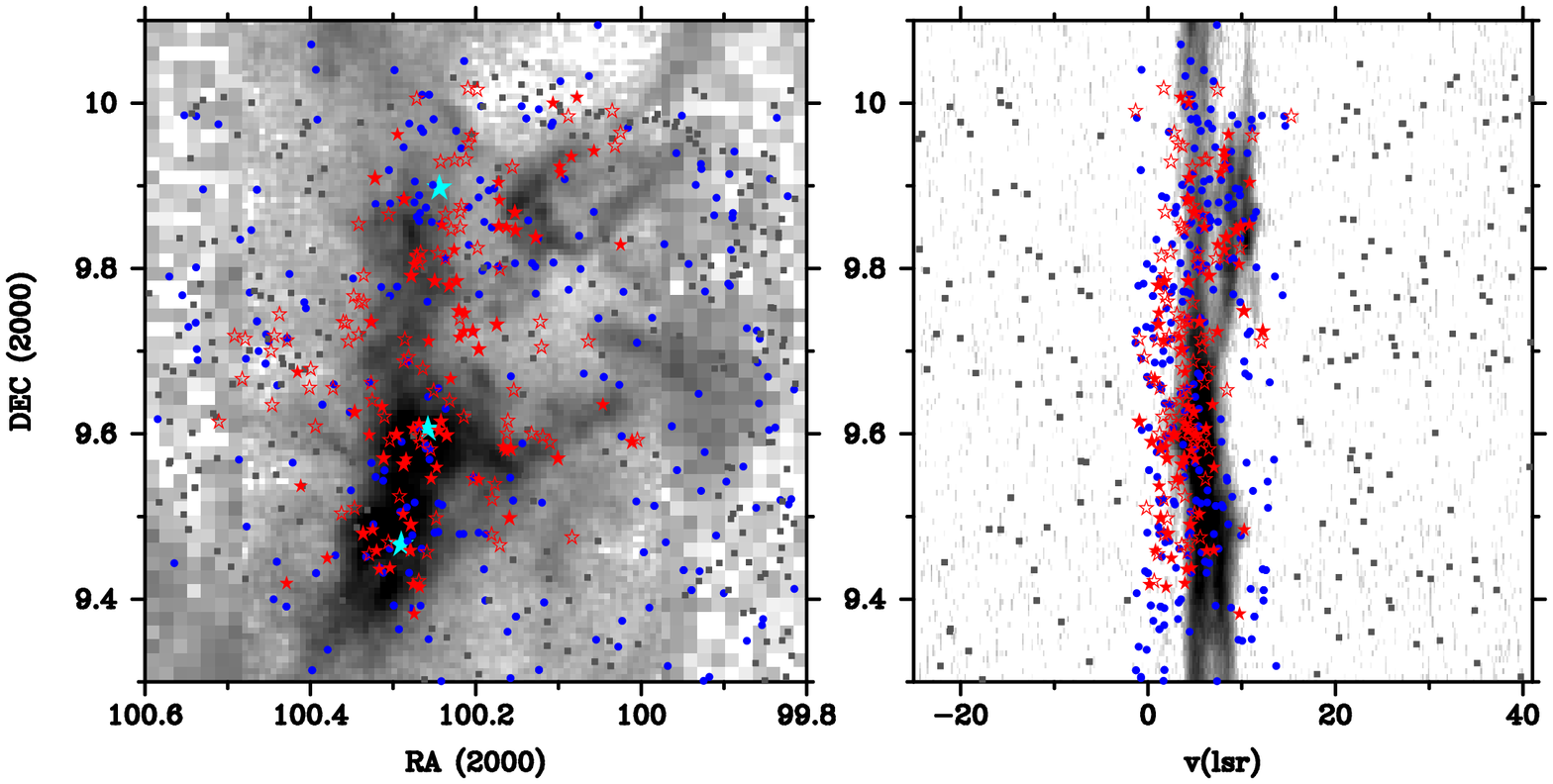}
\end{center}
\caption{NGC 2264 radial velocity targets superimposed on the $^{13}$CO integrated intensity
map (grayscale) from Ridge et al. (2006) (left panel) and the radial velocity
targets are overlaid on a position-velocity plot (right panel) summing over the RA range in the 
left panel. The lower resolution portion at the edges of the left panel is 
from J. Bally (unpublished). Star symbols represent members as determined from IRAC infrared excesses (filled stars) or
the detection of Li I absorption (open stars); blue circles denote non-excess stars that
were not observed for Li absorption, but whose velocities -2 \kms\ $< V(lsr) < 15$ \kms\ suggest
possible membership.  
Gray squares represent stars with LSR velocities $< -2$ and $>$ 15 \kms\  that
are unlikely to be cluster members. 
The non-members (gray squares) are roughly distributed uniformly 
in velocity and declination.
From north to south, the cyan stars in the left panel correspond to the position of S Mon, 
the Spokes Cluster center, and the Cone cluster center.}
\label{pv}
\end{figure}

Figure \ref{pv} shows an overview of the spatial and kinematic properties of our
sample and its correlation with the molecular gas in the region.
The left panel shows the positions of all the radial velocity targets
overplotted on the corresponding $^{13}$CO intensity map from
\citet{ridge2006} for comparison.  In the right panel, we show the
position-velocity (PV) distribution of the stars summed over right ascension
and plotted as function of declination to correspond roughly to the long axis of the cluster,
with the corresponding $^{13}$CO PV plot.
Targets that show infrared (IR) excess emission, as defined by 
$K_s$ - [3.6] $>$ 0.4 and [3.6] - [4.5] $>$ 0.2, where $K_s$ is from 2MASS and the other colors
are IRAC bands 1 and 2, are marked as filled stars; targets with detectable Li absorption
but no IR excess are shown as open stars.

The spatial and kinematic correlations between stellar members and molecular gas are evident.
The targets with projected spatial positions well off the main molecular gas 
emission are mostly taken with Hectochelle, chosen to use ``left-over'' fibers; these are
largely non-members, as expected, based on their radial velocities that
differ substantially from the cluster mean.  

\begin{figure}
\begin{center}
\includegraphics[scale=0.75, angle=-90]{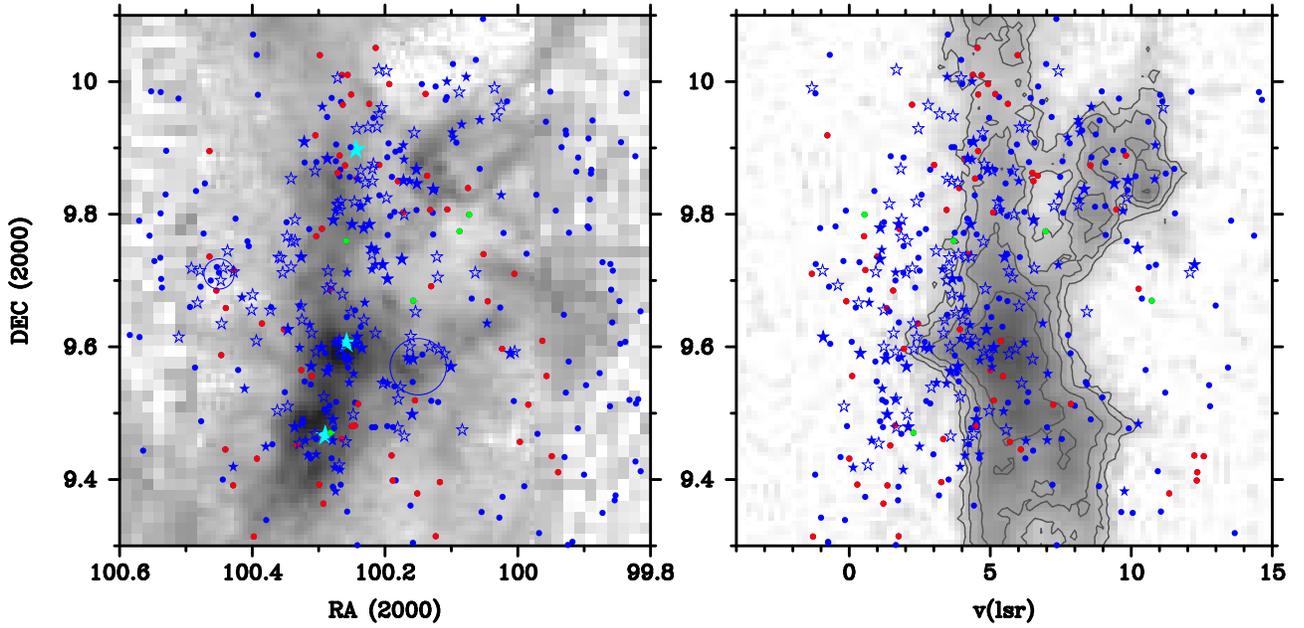}
\end{center}
\caption{Radial velocities of known and likely members again superimposed upon the $^{13}$CO integrated intensity
map, as in Figure 1 projected spatially (left) and in the declination position-velocity map (right).
Symbol shapes are the same as in Figure 1 - filled stars for infrared-excess stars, open stars for
Li absorption detections, and filled circles for possible members that do not have infrared-excess (or lack photometry) and/or
Li absorption measurements. The red circles are sources without an 
infrared-excess excess and do not have IMACS observations to detection
Li absorption. The green circles are sources without infrared-excess, have IMACS observations.
The large open circles highlight two regions that show a clustering of blue-shifted stars.
From north to south, the cyan stars in the left panel correspond to the position of S Mon, 
the Spokes Cluster center, and the Cone cluster center.
}
\label{pvzoom}
\end{figure}

Figure \ref{pvzoom} focuses on the same comparisons between stars and gas in more detail by
removing stars with LSR velocities $<$ - 2 \kms\ and $>$ 15 \kms.
The overall kinematic properties of the stars are similar to that found by \citet{furesz2006}
(compare the right panel of their Figure 6), but by approximately doubling the number of members with 
radial velocity measurements several aspects become much clearer.  First, the addition
of many new members in the region of the Spokes Cluster \citep{teixeira06} at
$\delta \sim 9.6^{\circ}$, RA $\sim 100.25^{\circ}$ shows a substantial velocity dispersion with an extension
to blueshifted velocities, corresponding to a similar feature in the $^{13}$CO emission 
between 2 \kms\ and 5 \kms. 
Second, the population of redshifted stars in the north corresponding to the molecular gas
component at $V(lsr) \sim$ 10 \kms\ is better defined; there appears to be a gap at
$\delta$ $\sim$ 9.85, V(lsr) $\sim$ 7 \kms\ in both distributions.  Finally, there is a ``blueshifted
population with velocities -2 \kms\ $\lesssim$ V(lsr) $\lesssim$ 2 \kms\ which does not have corresponding
$^{13}$CO emission and was not apparent in \citet{furesz2006}.

\begin{figure}
\begin{center}
\includegraphics[scale=0.8]{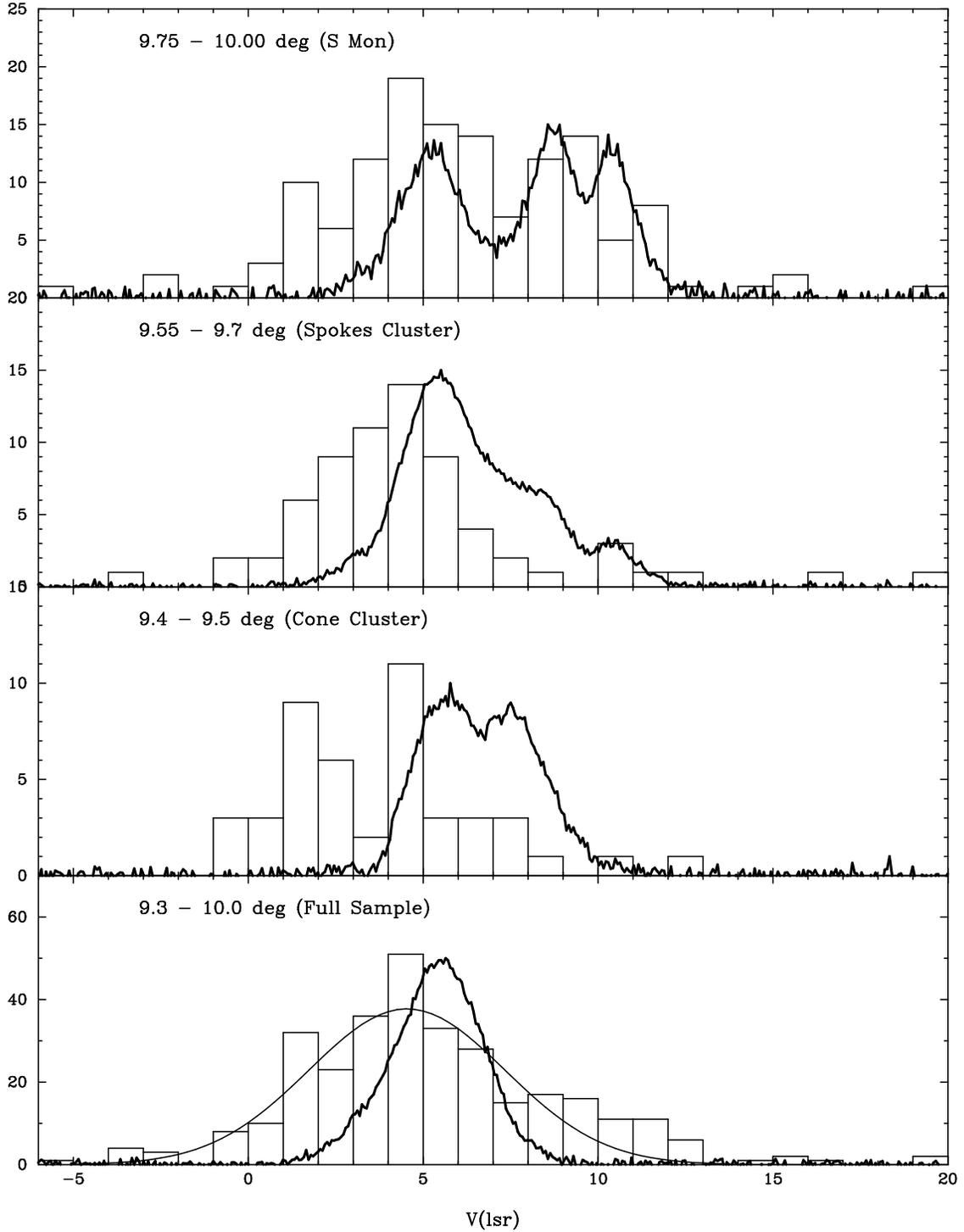}
\end{center}
\caption{RV histograms for stars and gas (thick line) for selected bins in declination. Note the systematic blue shift of the stellar
distribution compared to the gas in the southern part of the cluster. The smooth curve in the
bottom panel is a Gaussian fit to the stellar RV histogram, yielding a one-dimensional velocity dispersion of 
$\sigma$ = 2.8 \kms; after accounting for the average error of 1.2 \kms, the velocity dispersion is reduced to 2.5 \kms. Note that
we only plot stars with RA between 100.4 and 100.05 to focus on the main cluster area and limit contamination.}
\label{histos}
\end{figure}

To further examine the velocity distributions relative to the gas, we have plotted
histograms of the stellar and gas velocities in Figure \ref{histos} in the noted declination bins.
We only include a range of RA between 100.05$^{\circ}$ and 100.4$^{\circ}$
to focus on the main cluster and limit the contribution of the dispersed population.
The histograms show that while there is a strong overlap in velocity space between stars
and molecular gas, there is a modest but significant population of blueshifted stars which
do not have associated $^{13}$CO emission at their velocity.  
A substantial fraction of these blueshifted stars show either infrared excess or
strong Li I absorption (Figure \ref{pvzoom}), indicating that they are not substantially
older foreground objects.  Moreover, contamination by field stars
(for which we do not have either Li measurements or observed infrared excess) is unlikely to 
have a major effect on this distribution; we estimate a contamination of
$\sim$13 non-members in Figure \ref{histos}. We determine this number by counting the number of stars
with velocities between $-35$ \kms\ $<$ V(lsr) $<$ -5 \kms\ and 20 \kms\ $<$ V(lsr) $<$ 50 \kms\ 
in the same range of right ascension and assuming a uniform distribution in radial velocity.


The other obvious feature of note is the redshifted population of stars
at $9.8^{\circ} < \delta < 10^{\circ}$ (Figure \ref{pvzoom}). This group
corresponds in velocity space to a corresponding clump of $^{13}$CO emission
at 7 \kms\ $<$ V(lsr) $<$ 12 \kms.  There is a gap or minimum in the CO emission at
V(lsr) $\sim$ 7 \kms, and 
\textbf{
there is an indication of a similar minimum in the stellar 
velocity distribution, though it is statistically insignificant
}
(Figure \ref{histos}).  We suggest that this structure is the result
of the strong wind of S Mon driving a bubble into the molecular gas, triggering
star formation as the gas is compressed.

The histogram of the full sample with in a RA range of 100.05$^{\circ}$ and 100.4$^{\circ}$ and a Dec. range
of 9.35\degr\ to 10.0\degr\ is shown in the bottom panel of Figure \ref{histos}. We fit a Gaussian to this
histogram and derive a velocity dispersion of 2.8 \kms.  We account for the average velocity
error of 1.2 \kms\ for the stars used in the calculation by subtracting this error in quadrature from the
velocity dispersion and the corrected velocity dispersion is then 2.5 \kms.

\begin{figure}
\begin{center}
\includegraphics[scale=0.8]{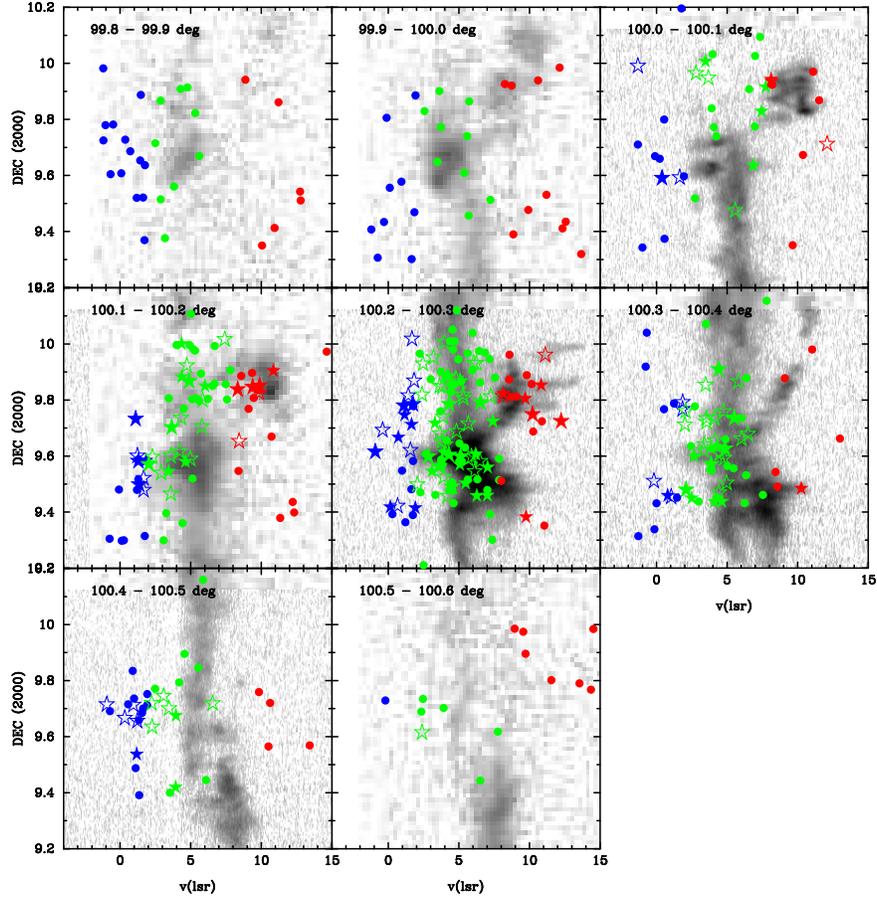}
\end{center}
\caption{PV plots for different bins in RA across the cluster. Important to note are the 100.2 to 100.4 bins which show that moving north
in the cluster there are three distinct velocity components. The blue points are used for stars with 
V (lsr) $\le$ 2 \kms, the green points are used for stars with
 2 \kms $<$ V(lsr) $<$ 8 \kms, and the red points are used for stars with V(lsr) $\ge$ 8 \kms. }
\label{racuts}

\end{figure}

Figure \ref{racuts} provides another view of the cluster structure in terms
of position-velocity maps, but by stepping across the cluster in 0.1\degr\ swaths of RA.  The
symbols have the same meaning as in Figure \ref{pvzoom}, but they are
now color-coded to emphasize blue- and red-shifted populations.  
Here we call attention to the small group of stars shown in the bottom
left panel (RA = 100.4 to 100.5)
centered around V(lsr) $\sim$ 2 \kms, $\delta \sim 9.5^{\circ}$ and
not associated with $^{13}$CO emission.  Only one of these stars exhibits
infrared excess, suggesting that this might be a slightly older group; on the other hand,
many show Li absorption. The population blueshifted stars is apparent
in nearly all RA bins across the cluster.

\section{Discussion}

\subsection{Kinematics and gravitational binding}

The overall kinematic structure of NGC 2264 bears a striking similarity to that found
in the Orion Nebula Cluster (ONC).  As shown in Figure 3 of \citet{tobin2009}, the position-velocity
plot of both stars and gas shows a northern region redshifted with respect to the central
region. The right panel of Figure \ref{pvzoom} shows qualitatively similar behavior, although more clearly
in the gas than in the stars.  One also observes the same asymmetry between the
stellar and gas velocities, with a population of stars blueshifted with respect to
the molecular gas.  The 1-dimensional velocity dispersions of the two clusters
are also similar, $\sigma$ = 3.1 \kms\ for Orion \citep{furesz2008} and $\sigma$ = 2.5 \kms\ for NGC 2264. 
NGC 2264 does differ from the ONC with its two components in the molecular gas
in the north near S Mon (three if the clump west of S Mon is considered), 
and the gas in the ONC has only one component.
The stars in the northern part of the ONC also have a smaller velocity dispersion with respect
to the center of the ONC, while in NGC 2264 the stars in the north have a larger spread in velocity than
the northern ONC stars.

\citet{proszkow2009} suggested that the transition from redshifted stars and gas in the
northern regions of the ONC to more blueshifted velocities near the cluster center could
be the signature of gravitationally-driven infall.  The interpretation is qualitatively
similar to using ``caustics'' in position-velocity space to infer
gravitational motion in galaxy clusters \citep[e.g.,][]{geller2013}.  To see whether a similar interpretation
could hold for NGC 2264, we compute a ``binding velocity'' $v_b \sim (2 GM/r)^{1/2}$.
\citet{crutcher1978} estimated a total gas mass of $\sim$8000 $\msun$
spanning a total length of $\sim$10 pc. Using this mass and setting $r \sim 5$~pc yields 
v$_b$ $\sim$ 3.6 \kms.

However, while this scenario was plausible for Orion, it seems less likely to
explain the red-shifted population in the north of NGC 2264. This group has 
V(lsr) $\sim$ 10 \kms\ which is redshifted by about
5 \kms\ with respect to the gas and stellar velocities at the position of the
Spokes Cluster.  Unless the gas mass (which is the dominant contributor) is underestimated
by roughly a factor of four, it seems very likely that this redshifted population is
not bound to the rest of the cloud.  Rather, we suggest this group is more likely to be the
result of blowout by the wind from S Mon; this would also explain
the ``hole'' in the gas at $\delta \sim 9.8^{\circ}$, $V(lsr) \sim 7$ \kms.

With respect to the rest of the cluster, the binding velocity calculated
from the molecular gas is larger than the 1-dimensional 
velocity dispersion ($\sigma_v$ = 2.5 \kms)
within a restricted range of RA (bottom right panel of Figure \ref{histos}). However, 
if the velocity dispersion is isotropic, the 3-dimensional velocity 
dispersion would be $\sim$4.3 \kms\ and thus suggests that
the region is unbound.
In addition, if our suggestion that a redshifted population is missing from our
optical sample is correct, the 1-d velocity dispersion of the total stellar population is larger than we measure here.
Resolution of these question will wait until we have proper motions from Gaia. 
For now, we conjecture that the true situation is a mix of possibilities, with some 
fraction of the stars that will eventually be left behind in a bound cluster and some dispersing.


\subsection{Blueshifted population}

\begin{figure}
\begin{center}
\epsscale{0.8}
\plottwo{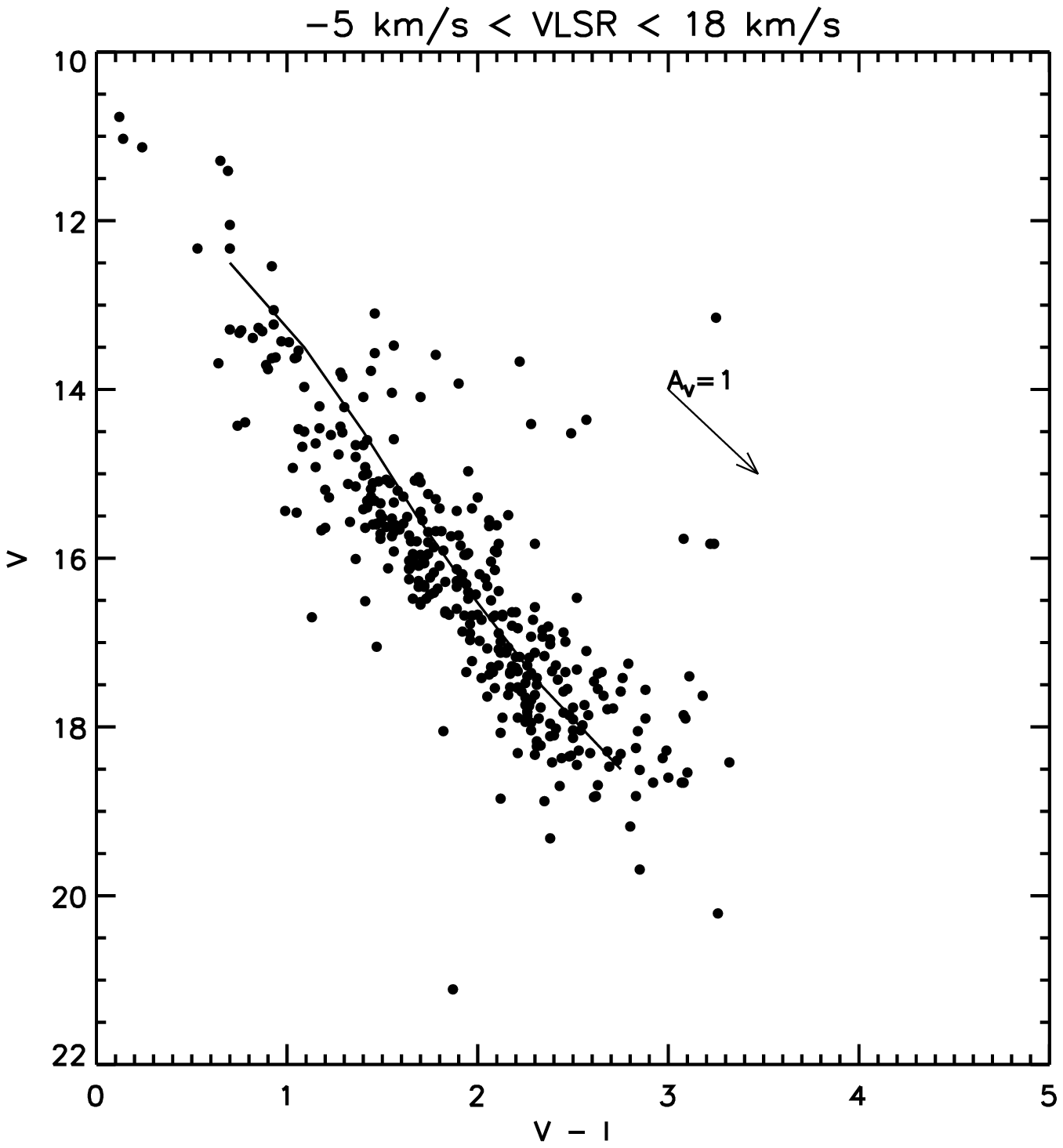}{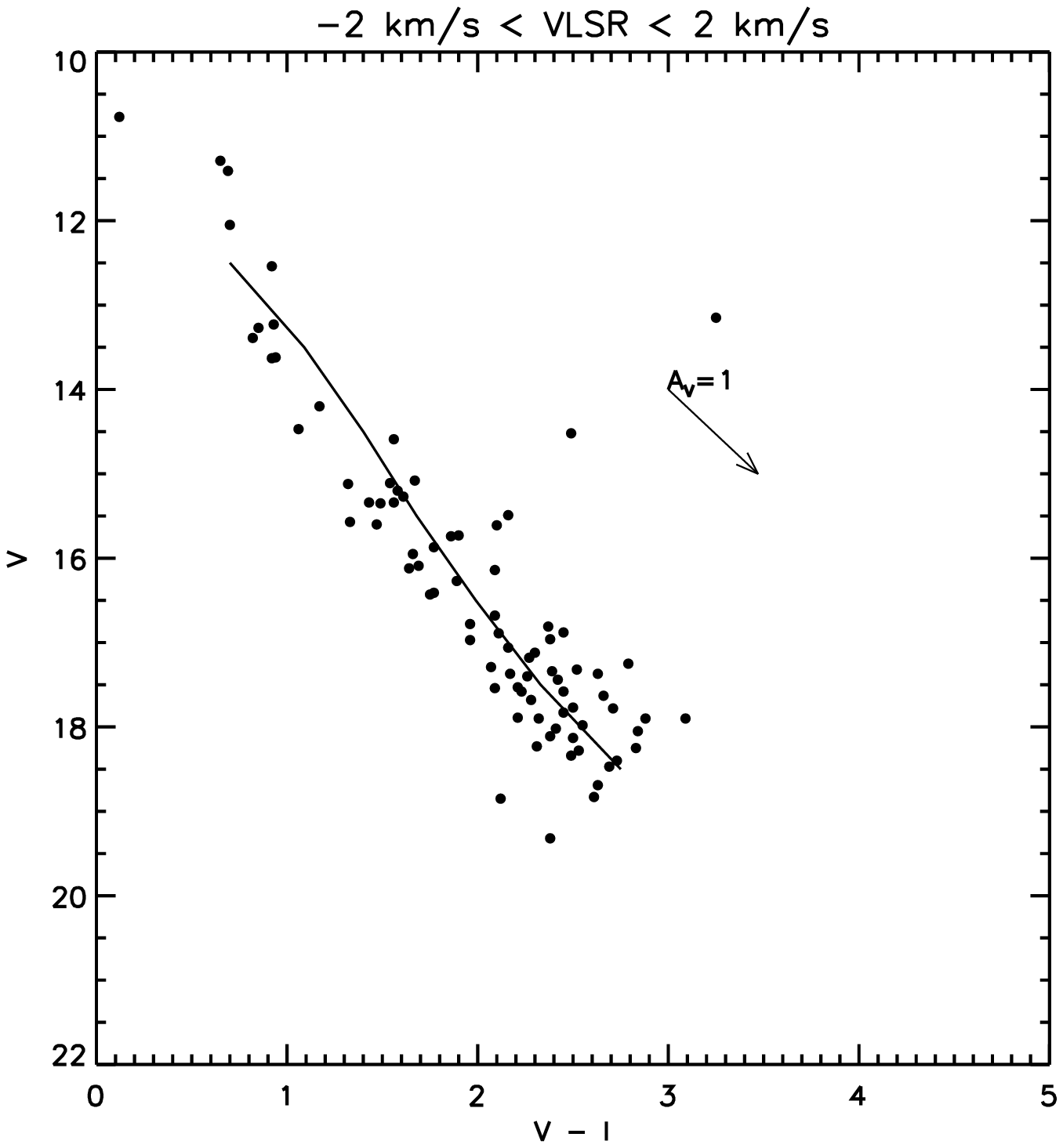}
\plottwo{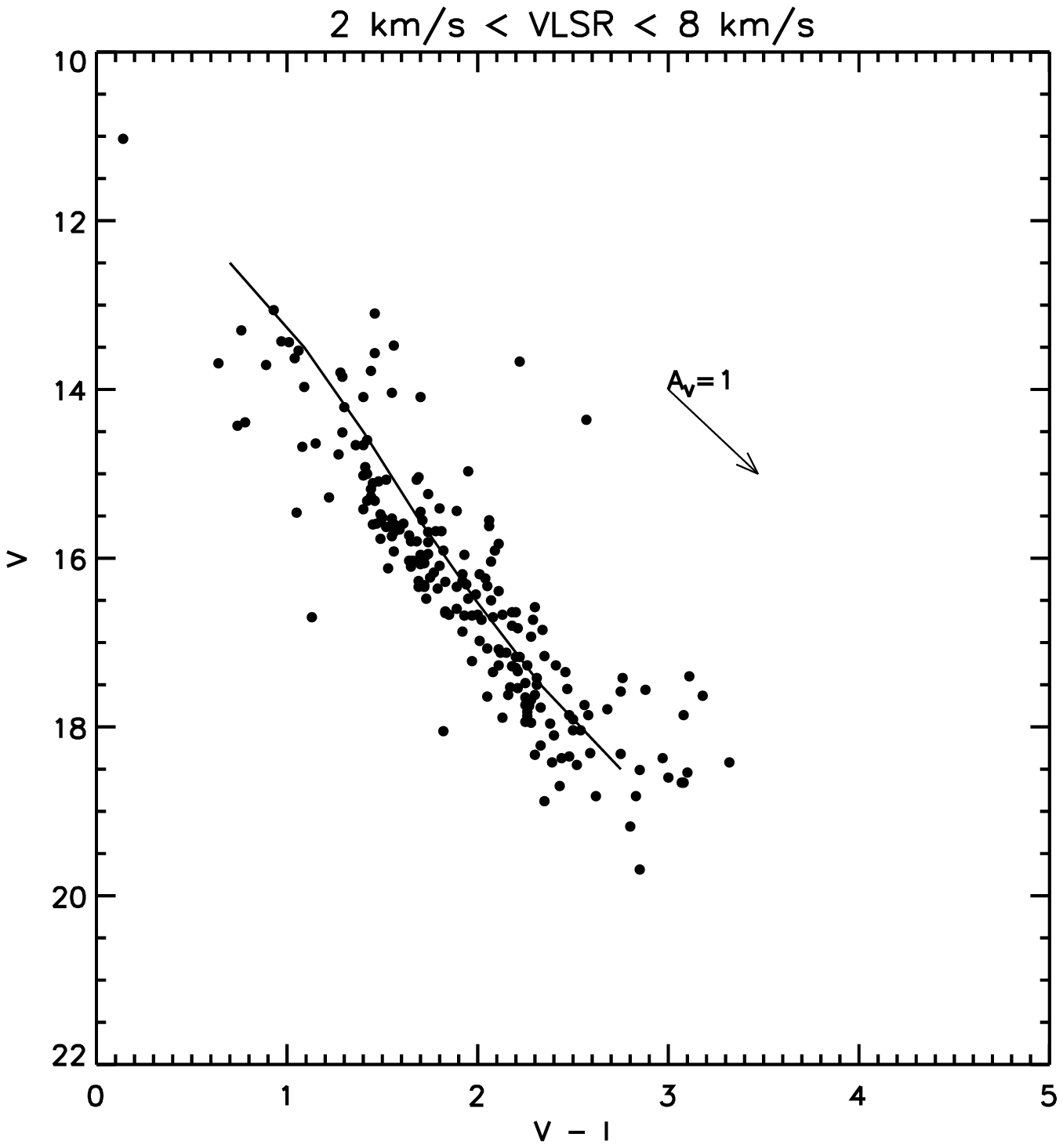}{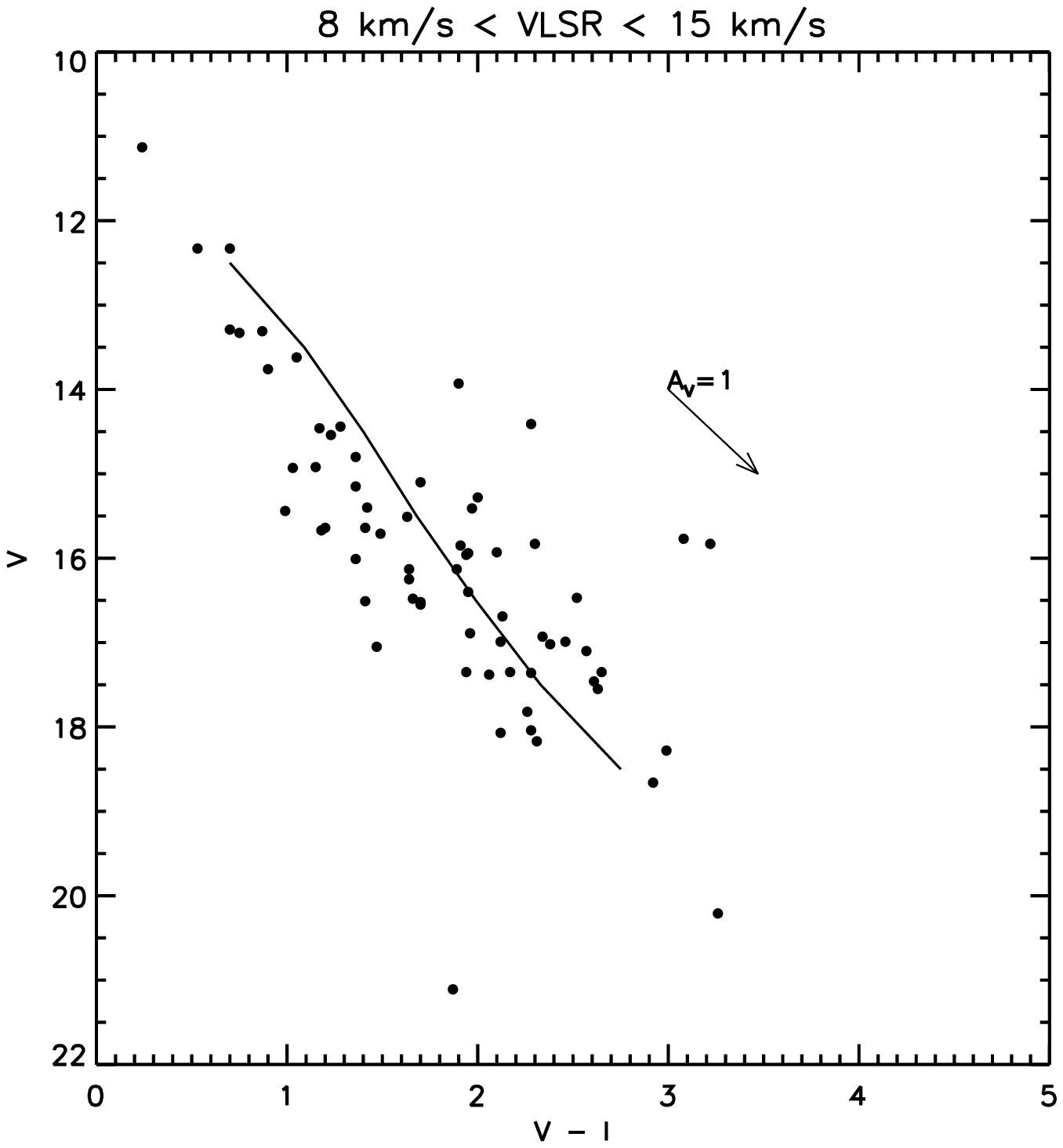}
\epsscale{0.35}
\plotone{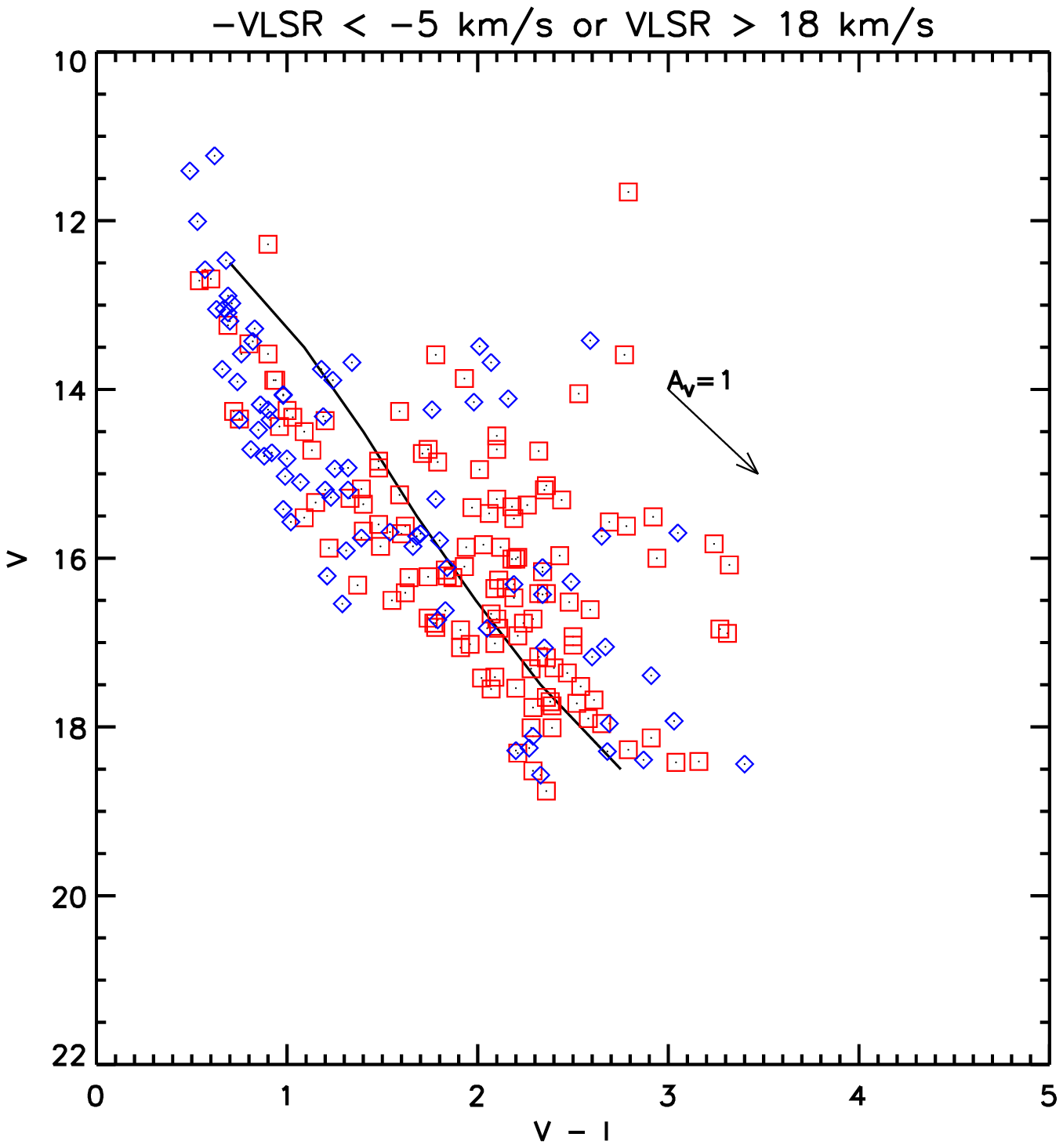}

\end{center}
\caption{\small Optical color magnitude diagrams of our NGC 2264 targets with optical photometry from Rebull et al. (2001);
the photometry are not corrected for extinction. The top left panel is
the CMD of stars with radial velocity that is consistent with cluster membership, -5.0 \kms\ $<$ V(lsr) $<$ 18.0 \kms.
The top right panel shows stars within the group that is blue-shifted of the gas velocity -2.0 \kms\  $<$ V(lsr) $<$ 2.0 \kms.
The middle left panel shows stars with the main cluster velocity, 2.0 \kms\  $<$ V(lsr) $<$ 8.0 \kms. The middle right
panel shows stars that a redshifted of the main cluster 8.0 \kms\  $<$ V(lsr) $<$ 15.0 \kms.
The bottom panel shows stars that are anti-correlated with the cluster velocity; the blue or red boxes
denote whether the stars is significantly blue or red-shifted away from the cluster velocity.
The \textit{solid line} in both panels is the median V-I color for stars with velocities consistent with NGC 2264.
The CMDs of the three groups within the cluster velocity are all consistent with each other, showing that they are at 
similar distances and ages. The stars outside the main cluster velocities show a significantly different CMD that is comprised
of both foreground and background sources.}
\label{optcmd}
\end{figure}

As discussed in the previous section, it appears that a substantial fraction of the blueshifted stars
are causally-related to NGC 2264 if not members, based on the indicators of youth -
infrared excesses and Li absorption.  One possibility for explaining why this population
is not at the same velocity as the molecular gas is that it might be older,
resulting from a previous episode of star formation.
To test this idea, we constructed color-magnitude diagrams for the various kinematic groups.  As seen in 
Figure \ref{optcmd}, the color magnitude diagram (CMD) of the ``field'' stars with velocities 
$<$ -5 \kms\ and $>$ 18 \kms\ from the cluster mean
shows a much larger dispersion around the median color-magnitude line of the cluster.
We observed an excess of faint stars, as would be expected for foreground main sequence stars,
and brighter, redder background giants. Moreover, the blue-shifted stars tend to be brighter and bluer in color,
while the red-shifted stars are fainter, redder, and have more dispersion in magnitude.

The CMDs of the various probable cluster member velocity groups 
are essentially indistinguishable except for the redshifted group 
with 8 \kms\ $< V(lsr) < 15$ \kms, which exhibits a broader
dispersion. This could be due to contamination 
\textbf{
because the northern region, where
much of the 8 \kms\ $< V(lsr) < 15$ \kms\ population is found,
}
has less extinction to cut out
background stars (and is also sparser).
The blue-shifted group  (-2 \kms\ $< V(lsr) < 2$ \kms)
appears to have a larger fraction of redder stars than the main cluster as well.
However, because these photometric data have not been de-reddened, it is
impossible to decouple evolutionary effects or population differences from extinction. 
Notice that the reddening vector shown in Figure \ref{optcmd} is nearly parallel to the median
CMD of the cluster. On the whole, there is a significant spread
in $V$ at a given $V-I_C$ color, but much of this is undoubtedly
due to an unresolved binary population; other sources of error are probably
present but poorly-understood \citep{jeffries12}. In any case, there is little evidence
for any systematic age differences larger than about 
$\sim 2-3$~Myr, comparable to the median estimated age of the cluster \citep{sung2010}. 

Systematic ejection of stars toward Earth seems unlikely if not implausible, especially
because the ONC shows the same excess of blueshifted stellar velocities relative to the
molecular gas velocities.  We next consider other possibilities for the asymmetry between 
the stellar and gas velocities. 
The first possibility recognizes that we have an optically-biased sample of
radial velocities; stars moving away from us would move more
deeply into the molecular cloud and thus have much higher line-of-sight extinctions.
It seems very likely that some bias exists; as shown in Figure \ref{histos},
we observe relatively few stars at the reddest velocities of the molecular gas,
which would be surprising if not attributable to sample bias.  Whether there
exists a substantial population of redshifted stars with velocity shifts of comparable
magnitude to the blueshifted population is not clear.  If such a population did exist,
the overall velocity dispersion would increase to $\sigma_v \sim 5$ \kms, making it
even less likely that any part of the region is gravitationally-bound.

An alternative possibility to explain the asymmetry between stars and gas is preferential dispersal
of blueshifted (nearside) gas by stellar energy input.  We argued above that S Mon is blowing
away gas, producing the ``hole'' in the $^{13}$CO emission at $\sim 7$ \kms, $\delta \sim 9.8^{\circ}$. 
It seems conceivable that this O7 star could have dispersed additional molecular gas.
\citet{schwartz85} argued that S Mon is responsible for photoionizing the rim of the
Cone Nebula at a projected distance of about 6 pc.  
Furthermore, they argued that S Mon and the Cone Nebula might actually
lie a substantial distance in the foreground from other dense molecular gas in the
region (basically, the Spokes Cluster region).  Whether or not this picture is correct,
it is possible that S Mon photoevaporated and photoionized the gas originally on
the near side of the original molecular cloud. 

 The simulations of
photodissociation and ionization of initially molecular gas with magnetic fields
and turbulence by \citet{arthur11} show substantial evacuation of a region 
of size 4 pc and initial density $n = 10^3 {\rm cm^{-3}}$ in only $4 \times 10^5$~yr
for an assumed O9 star, which emits a factor of $3-4$ fewer ionizing photons than
an O7 star is expected to provide \citep[][Table 2.3]{osterbrock2006}.
Thus, it is possible that there had been and episode of star formation in the near side of the
molecular cloud, prior to the formation of S Mon. Once S Mon formed and began photoionizing
the molecular gas associated with the previous generation of star formation, the molecular gas
removal caused the previously formed stars to become unbound and expand \citep[e.g.,][]{ladalada2003}.
This scenario would produce a population of stars systematically blue-shifted with respect to the cloud velocity,
because we are able to observe the blue-shifted stars due to their low extinction, but the red-shifted
stars move into the molecular cloud and become too extincted to determine radial velocities in the optical.
We therefore suggest that the observed kinematic differences between the stellar population
and the molecular gas result from a combination of optical bias against stars moving
into the molecular gas and blowout of molecular gas on the near side by S Mon.
The low extinction
observed toward some members \citep{walker1956,park2000} is clear evidence for some gas dispersal to have taken place.

While we cannot determine a systematic age difference between the blue-shifted population
and the rest of the cluster, \citet{sung2010} argued that the `field' and `halo' populations
of NGC 2264 are older than the rest of the cluster. These regions are designated as such because
they they are spread across the NGC 2264 region on the sky and the blue-shifted population of
stars is found at all positions in NGC 2264 (see Figure \ref{racuts}). It is noteworthy that Orion
also has several know foreground populations as well, projected on the same region of the sky
as the ONC \citep[e.g.,][]{bally2008,alves2012}, possibly contributing to the blue-shifted population observed there.

\section{Summary}

The comparison of the spatial-kinematic properties of stars and molecular gas
suggests that NGC 2264 may be better understood as a loose collection
of star-forming clumps rather than as a single, distinct, strongly-bound cluster.
Future astrometric surveys to determine distances and proper motions are needed
to fully develop the true picture of this object.

We wish to thank the anonymous referee for comments which improved the clarity
of the manuscript.
The authors wish to thank the staff of the MMT and Magellan telescopes
and the Hectochelle queue observers of 2007, 2008, 2009, and 2010 for their efforts in
obtaining the data used in this paper.
We also thank M. Walker and E. Olszewski for
assistance with MIKE Fibers data acquisition/reduction and 
Jesus Hernandez for useful discussions. J. T. and L. H. acknowledge funding from
NSF grant AST 08070305. J.J.T. acknowledges support provided by NASA through Hubble Fellowship 
grant \#HST-HF-51300.01-A awarded by the Space Telescope Science Institute, which is 
operated by the Association of Universities for Research in Astronomy, 
Inc., for NASA, under contract NAS 5-26555. This research has made use of NASA's Astrophysics Data System.
 The National Radio Astronomy 
Observatory is a facility of the National Science Foundation 
operated under cooperative agreement by Associated Universities, Inc.

{\it Facilities:} \facility{Spitzer (IRAC)}, \facility{Magellan:Clay (MIKE)}, \facility{MMT (Hectochelle)}
\begin{small}
\bibliographystyle{apj}
\bibliography{ms}
\end{small}

\clearpage
\begin{deluxetable}{llllllllllll}
\rotate
\tablewidth{0pt}

\tabletypesize{\tiny}
\tablecaption{Hectochelle Observations}
\tablehead{
  \colhead{Field ID} & \colhead{Date} & \colhead{Julian Date} &  \colhead{RA} & \colhead{Dec.} & \colhead{Airmass} & \colhead{Exposure Time} & \colhead{Binning}  & \colhead{Filter} & \colhead{Zeropoint Shift\tablenotemark{a}} & \colhead{Number of Targets} & \colhead{Seeing}\\
                 & \colhead{UT Date} & \colhead{(2400000)} &  \colhead{(J2000)} & \colhead{(J2000)} & & \colhead{\# $\times$ seconds} & & & \colhead{km s$^{-1}$} & & \colhead{arcsec} }
\startdata
F1-E1-2007 & 10-12-2007 & 54385.42 & 06:40:57.74  & 09:40:26.20 & 1.33 & 3x1200 & 2x2 & RV31 & -0.18 $\pm$ 0.02 & 147 & 1.07\\
F1-E2-2007 & 10-26-2007 & 54399.42 & 06:40:57.74  & 09:40:26.20 & 1.19 & 3x1200 & 2x2 & RV31 & +0.23 $\pm$ 0.02 & 147 & 0.64 \\
F2-E1-2007 & 10-12-2007 & 54385.48 & 06:40:39.15  & 09:45:23.48 & 1.12 & 3x1200 & 2x2 & RV31 & -0.08 $\pm$ 0.04 & 147 & 0.61 \\
F2-E2-2007 & 10-28-2007 & 54401.40 & 06:40:39.15  & 09:45:23.48 & 1.25 & 3x1200 & 2x2 & RV31 & +0.23 $\pm$ 0.03  & 147 & 1.4 \\
F3-E1-2007 & 10-29-2007 & 54402.44 & 06:41:06.09  & 09:42:34.69 & 1.12 & 3x1200 & 2x2 & RV31 & -0.02 $\pm$ 0.02  & 144 & 1.97\\
F3-E2-2007 & 11-25-2007 & 54429.40 & 06:41:06.09  & 09:42:34.69 & 1.08 & 3x1200 & 2x2 & RV31 & -0.18 $\pm$ 0.08  & 144 & 2.32\\
F4-E1-2007 & 10-29-2007 & 54402.49 & 06:40:51.25  & 09:33:13.59 & 1.08 & 1x1200 & 2x2 & RV31 & +0.14 $\pm$ 0.04  & 143 & 1.97\\
\\
\enddata
\tablenotetext{a}{Shift applied is relative to the velocities in \citet{furesz2006}.}
\end{deluxetable}

\begin{deluxetable}{lllllllllll}
\rotate
\tablewidth{0pt}

\tabletypesize{\scriptsize}
\tablecaption{MIKE Fibers Observations}
\tablehead{
  \colhead{Field ID} & \colhead{Start Date} & \colhead{Julian Date} &  \colhead{RA} & \colhead{Dec.} & \colhead{Airmass} & \colhead{Exposure Time} & \colhead{Binning}  & \colhead{Filter} & \colhead{Zeropoint Shift} & \colhead{Number of Targets}\\
                 & \colhead{UT Date} & \colhead{(2400000)} &  \colhead{(J2000)} & \colhead{(J2000)} & & \colhead{\# $\times$ seconds} & & & \colhead{km s$^{-1}$} }
\startdata
A-1 & 12-10-2009 & 55175.81 & 06:41:00.98  &  09:32:44.4   & 1.4 & 3x1200 & 2x2 & Mg &  0.05$\pm$ 0.1 & 220\\
B-1 & 12-11-2009 & 55179.81 & 06:40:48.90  &  09:51:44.4  & 1.4 & 3x1200 & 2x2 & Mg &  -0.13 $\pm$ 0.05 & 215\\
A-2 & 12-14-2009 & 55176.80 & 06:41:00.98  &  09:32:44.4  & 1.4 & 3x1200 & 2x2 & Mg & 0.31$\pm$ 0.05 & 220\\
B-2 & 12-15-2009 & 55180.83 & 06:40:48.90  &  09:51:44.4  & 1.5 & 3x900 & 2x2 & Mg & 0.08 $\pm$ 0.0.04 & 215\\

\enddata
\tablenotetext{a}{Shift applied is relative to the velocities in \citet{furesz2006}.}
\tablecomments{} 
\end{deluxetable}

\begin{deluxetable}{lllllllll}
\tablewidth{0pt}
\tabletypesize{\scriptsize}
\tablecaption{Velocity Summary}
\tablehead{
  \colhead{2massID} & \colhead{RA} & \colhead{Dec.} &  \colhead{$\overline{RV}$\tablenotemark{a}} & \colhead{N Epochs\tablenotemark{b}}  & \colhead{Lithium EW} & \colhead{Spectral Type} & \colhead{K - [3.6]} & \colhead{[3.6] - [4.5]}\\
  & \colhead{(J2000)} & \colhead{(J2000)} & \colhead{(km s$^{-1}$)} &  & \colhead{(\AA)} & \\

   }
\startdata
0639037+094023 & 06:39: 3.75 & 09:40:23.4 & 5.32 $\pm$ 0.13 & 4 & -99.00 & -99.0 & -99.00 & -99.00 \\
0639152+093937 & 06:39:15.25 & 09:39:37.8 & 29.88 $\pm$ 0.24 & 7 & -99.00 & -99.0 & -99.00 & -99.00 \\
0639153+095303 & 06:39:15.37 & 09:53:03.6 & 102.33 $\pm$ 0.95 & 3 & -99.00 & -99.0 & -99.00 & -99.00 \\
0639155+092444 & 06:39:15.55 & 09:24:44.8 & 10.94 $\pm$ 0.07 & 2 & -99.00 & -99.0 & -99.00 & -99.00 \\
0639156+093912 & 06:39:15.63 & 09:39:13.0 & 1.43 $\pm$ 0.42 & 9 & -99.00 & -99.0 & -99.00 & -99.00 \\
0639157+092929 & 06:39:15.70 & 09:29:29.4 & 18.65 $\pm$ 0.02 & 3 & -99.00 & -99.0 & -99.00 & -99.00 \\
0639163+091953 & 06:39:16.37 & 09:19:53.3 & 50.59 $\pm$ 0.05 & 3 & -99.00 & -99.0 & -99.00 & -99.00 \\
0639164+093116 & 06:39:16.45 & 09:31:16.1 & 1.63 $\pm$ 0.11 & 2 & -99.00 & -99.0 & -99.00 & -99.00 \\
\enddata
\tablecomments{Abridged table of target IDs, positions, average radial velocities, Lithium 6707 \AA\ Equivalent Widths, Spectral Types, and near infrared colors.
Entries with -99 signify that these data are unavailable for a given target.}
\tablenotetext{a}{Velocities from \citet{furesz2006} are included in average. For a measurement to be included in the average R must be $>$ 6.0; R is defined in Equation 1.}
\tablenotetext{b}{Number of observations with R $>$ 6.0.}
\end{deluxetable}

\end{document}